\documentclass[aps,prl,showpacs,superscriptaddress,twocolumn]{revtex4}
\usepackage{amsmath}
\usepackage{amsfonts}
\usepackage{graphicx}
\usepackage{psfrag}
\usepackage{hyperref}
\usepackage{color}
\usepackage{siunitx}

\begin{document}

\title{Interference-induced phenomena in high-order harmonic generation from bulk solids}

\author{Viktor Szaszk\'{o}-Bog\'{a}r}
\email{viktor.szaszko-bogar@eli-alps.hu}
\affiliation{ELI-ALPS, ELI-HU Non-Profit Ltd., Dugonics t\'{e}r 13, H-6720 Szeged, Hungary}
\affiliation{Department of Optics and Quantum Electronics, University of Szeged, D\'{o}m t\'{e}r 9, H-6720 Szeged, Hungary}
\author{Katalin Varj\'{u}}
\affiliation{ELI-ALPS, ELI-HU Non-Profit Ltd., Dugonics t\'{e}r 13, H-6720 Szeged, Hungary}
\affiliation{Department of Optics and Quantum Electronics, University of Szeged, D\'{o}m t\'{e}r 9, H-6720 Szeged, Hungary}
\author{P\'{e}ter F\"{o}ldi}
\affiliation{ELI-ALPS, ELI-HU Non-Profit Ltd., Dugonics t\'{e}r 13, H-6720 Szeged, Hungary}
\affiliation{Department of Theoretical Physics, University of Szeged, Tisza Lajos k\"{o}r%
\'{u}t 84, H-6720 Szeged, Hungary}

\begin{abstract}
We consider a quantum mechanical model for the high-order harmonic generation in bulk solids. The bandgap is assumed to be considerably larger than the exciting photon energy. Using dipole approximation, the dynamical equations for different initial Bloch states are decoupled in the velocity gauge. Although there is no quantum mechanical interference between the time evolution of different initial states, the complete harmonic radiation results from the interference of fields emitted by all the initial (valence band) states. In particular, this interference is shown to be responsible for the suppression of the even order harmonics. The number of the observable harmonics (essentially the cutoff) is also determined by electromagnetic field interference.
\end{abstract}


\maketitle

\section{Introduction}
The first observation of photoemission spectra with clear high-order harmonics was reported using gaseous target media \cite{McPherson87,Ferray88}. In noble gas samples, the atoms driven by intense laser fields can be sources of secondary radiation ranging from the visible to the extreme ultraviolet regime. Besides the wide bandwidth, this radiation has excellent temporal and spatial coherence. Additionally, the  process of high-order harmonic generation (HHG) can be applied for the creation of attosecond pulses \cite{FT1992,HKSRMBCHDK2001,PTBMABMA2001}, which are ideal tools for time-resolved investigation of electronic processes. The first experiments that successfully generated such short pulses, triggered a fast development of measurement and control techniques as well as that of theoretical models \cite{DHKTSRCK2001,FFHFMSPEW2005,DSLMIVC2006,SIKV2006,CrKr2007,EPCSDMBK2008,KrSt2014,CSSVN2016}.

Typical intensities of the HHG signals are $5-10$ orders of magnitude lower than that of the applied exciting field, which is partially due to the low particle density in gas samples. One possible way to increase the conversion efficiency is using solid targets \cite{Marangos2011}. For the case of surfaces, there are two mechanisms the appearance of which strongly depends on the laser intensity: coherent wake emission (CWE)~\cite{QTMDMGA2006} and relativistic oscillating mirror (ROM)~\cite{Vincenti2014}. Besides surfaces, bulk solids
has also been demonstrated to produce high-order harmonics, e.g.~using few-cycle, mid infrared $(0.34-0.38\,\operatorname{eV})$ pulses impinging on a wide-bandgap zinc oxide (ZnO) crystal \cite{GDiCSADiMR11,GDSSzADR11,GDiCSNSzMADiMR12,GNDiCSSADiMR14}. Gallium selenide (GaSe) single crystal has also been used as a target when the control of terahertz high harmonic generation was studied \cite{SHLULHGMKKH14}. Considering real-time observation of crystal electrons, a non-perturbative quantum interference was identified between interband transitions as a salient HH generation mechanism \cite{HLScKHKKH15}. In Ref.~\cite{NGWBScGR16}, a direct comparison of the HHG signal was given in the solid and gas phases of argon and krypton. Properties of the HHG radiation itself have also been investigated in various samples \cite{GZSRYZPAKLK2010,SRPSWGPBPYNL14,You2017,Garg2016,Liu2017,Han2016}.

Besides the experimental results, several theoretical methods were developed in order to describe the interaction of intense electromagnetic field and matter. The quasi-classical method to explain high-order harmonics is the well known three step model \cite{Cor93}. One of the most successful quantum-mechanical treatments of the process was introduced by Lewenstein et al.~\cite{Lew94}, known as the strong field approximation model of HHG, which has also been applied for solid samples \cite{DRKS2011,HSH14}. Saddle-point approximation can also be used to evaluate the integrals that appear in the description of both gases and solids \cite{Sansone04,VMcOKCB14,VMcFB15}.

As a different approach, ab initio calculations can provide effective methods for solving the time-dependent quantum mechanical problem of driven many-electron systems. E.g., the nonlinear response to strong laser fields was studied by the three-dimensional time-dependent two-particle reduced-density-matrix theory (TD-2RDM) \cite{Brezinova17}. In Ref.~\cite{TDMKR17Nat}, time-dependent density functional theory (TDDFT) was used to investigate the dependence of the process on the ellipticity of laser field. Other TDDFT simulations revealed how the inhomogeneity of the electron-nuclei potential affects the process of HHG in solids \cite{TDMKR17}.

\bigskip

In the current work, we focus on the case of bulk solids. As we shall see, a one-dimensional model \cite{FP2017} for the crystalline solid with a bandgap having the order of a few $\operatorname{eV}$ can reproduce the qualitative features of the HHG spectra (plateau region and cutoff, including its intensity dependence). Working in single electron picture, we determine the Bloch states and the corresponding energies. The resulting band structure has a bandgap (between the valence band (VB) and the first conduction band (CB)) that is equal to that of the ZnO crystal. We solve the corresponding dynamical equations numerically and calculate the HHG signal using the expectation value of the dipole moment operator. Since in velocity gauge the time evolution of different Bloch states is not coupled, this calculation can be carried out for different VB initial states independently. Our aim is to show that despite this independence, the net HHG signal is strongly influenced by the interference of the radiation emitted by different initial states. Note that although spatial interference during propagation \cite{TTNM2003,KBHTV12} can also have significant effect on the HHG spectra, here we focus on time domain interference that can be detected at a fixed spatial point.

Our paper is organized as follows: first we discuss the physical and numerical aspects of the model to be used (Sec.~\ref{modelsec}). Then, in Sec.~\ref{resultsec}, the main properties of the calculated HHG spectra are analyzed: a general description (Subsec.~\ref{resultsec}\ref{spectra}) is followed by the physical reasons for  the suppression of the even harmonics (Subsec.~\ref{resultsec}\ref{evenodd}) and the position of the cutoff (Subsec.~\ref{resultsec}\ref{cutoff}).  Finally, we summarize our results in Sec.~\ref{summ}.

\section{Model and method}
\label{modelsec}
\subsection{Time-dependent Hamiltonian}

The single-electron Hamiltonian in a crystal is given by
\begin{equation}
H(t)=\frac{1}{2m}(\mathbf{p}-e\mathbf{A})^{2}
+U(\mathbf{r})+e\Phi, \label{Ham}
\end{equation}
where $e$ denotes the elementary charge and the lattice-periodic potential $U(\mathbf{r})$ represents the solid. The external electromagnetic field is taken into account via the scalar and vector potentials, $\Phi$ and $\mathbf{A}.$ For a given $U,$ one can look for the solution of the field-free problem in the form of Bloch states, $\Psi
_{n,\mathbf{k}}(\mathbf{r})=\exp(i\mathbf{kr})u_{n,\mathbf{k}}(\mathbf{r}%
)/\sqrt{\mathcal{V}},$ where $n$ denotes the band
index, $u_{n,\mathbf{k}}(\mathbf{r})$ are lattice-periodic functions and $\mathcal{V}$ is the crystal volume. The eigenvalue equation for these states reads
\begin{equation}
-\frac{\hbar^2}{2m}\left(\nabla+i\mathbf{k}\right)^2 u_{n,\mathbf{k}}(\mathbf{r})+U(\mathbf{r})u_{n,\mathbf{k}}(\mathbf{r})=E_n(\mathbf{k})u_{n,\mathbf{k}}(\mathbf{r}).
\label{keigen}
\end{equation}
As we can see, these equations for different $\mathbf{k}$ vectors are independent, thus, due to the periodicity of the functions $u_{n,\mathbf{k}}(\mathbf{r}),$ it is sufficient to determine the $\mathbf{k}$ dependent eigenenergies $E_n(\mathbf{k})$ in a single unit cell for all values of $\mathbf{k}$ separately. Having obtained the Bloch states and the eigenenergies $E_n(\mathbf{k})$ (i.e., the band scheme), we have the scene on which laser induced dynamics take place.

In actual numerical calculations one has to determine the electromagnetic gauge to be used. In the following, we consider velocity gauge, where the electromagnetic potentials can be expressed using the electric field
$\mathbf{F}(t)$ as $\mathbf{A}(t)=-\int_{-\infty}^{t}\mathbf{F}%
(t^{\prime})dt^{\prime},\Phi=0$. Realistically, we assume that the laser spot size is much larger than the unit cell, therefore the spatial variation of the laser field can be neglected. Within this dipole approximation, the Hamiltonian (\ref{Ham}) is diagonal in $\mathbf{k},$ thus the time evolutions of different Bloch-states $\Psi
_{n,\mathbf{k}}$ and $\Psi_{n,\mathbf{k}'}$ (with $\mathbf{k}\neq\mathbf{k}'$) as initial states are independent. However, all these states interact with the \emph{same} HHG modes, in other words, the fields emitted by the different initial states do interfere. This is the point we will analyze in Sec.~\ref{resultsec}.

\begin{figure}[tbh]
\includegraphics[width=8.5cm]{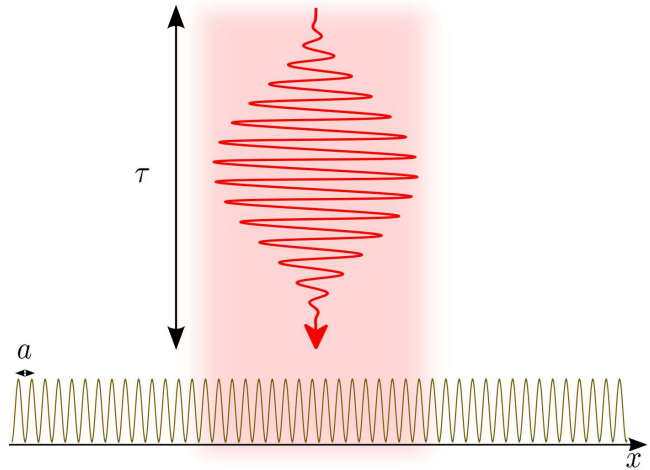}
\caption{Schematic view of an external laser driven periodic structure. The time duration of the pulse is denoted by $\tau$ and the lattice constant of the 1D crystal is given by $a$. Note that usually the interaction area is much larger than $a,$ thus dipole approximation can be used.}
\label{Fig1}
\end{figure}

\subsection{The source of the SHHG signal}
In order to solve the dynamics induced by the Hamiltonian (\ref{Ham}), the initial state has to specified. In other words, by expanding the quantum mechanical state $|\Phi(t)\rangle=\Phi(\mathbf{r},t)$ of the electron in terms of the Bloch states $|n,\mathbf{k}\rangle=\Psi
_{n,\mathbf{k}}(\mathbf{r})$ as $|\Phi(t)\rangle=\sum_{n,\mathbf{k}} c_{n,\mathbf{k}} (t)|n,\mathbf{k}\rangle,$ the coefficients $c_{n,\mathbf{k}}(t=0)$ should be given. (Note that for the sake of simplicity we assume periodic boundary conditions, and consequently the k-space will not be continuous, there will be a discrete, densely spaced series of k vectors.)

However, the most plausible assumption for the initial state of the crystal is that it is in thermal equilibrium before the arrival of the laser pulse. Thermal states are not pure quantum mechanical states, thus instead of a state vector we have to use a density matrix $\rho.$ Since the Bloch states are energy eigenstates of the crystal, all density matrices that represent thermal equilibria are diagonal in this basis. The corresponding statistical weights at room temperature and a band gap around $3\operatorname{eV}$ are practically nonzero only for the valence bands. That is, if we take a single VB into account with the corresponding band index $n_{0},$ we can write
\begin{equation}
\rho(t=0)=\frac{1}{N}\sum_{\mathbf{k}}|n_0,\mathbf{k}\rangle\langle n_0,\mathbf{k}|,
\label{rho0}
\end{equation}
where the constant $N$ equals the number of $\mathbf{k}$ vectors in the first Brillouin zone, and provides normalization: $\mathrm{Tr}\rho(0)=1.$ The time evolution of the density operator is governed by the von Neumann equation supplemented by a phenomenological term that takes unavoidable decoherence effects into account:
\begin{equation}
\frac{\partial}{\partial t}\rho(t)=-\frac{i}{\hbar}\left[H(t), \rho(t) \right] + \left. \frac{\partial}{\partial t}\rho(t)\right|_{dec},
\label{vN}
\end{equation}
where the second term on the right hand side forces the population towards the valence band with a rate of $\gamma_d$, and destroys the off-diagonal matrix elements (quantum mechanical coherences) with a rate of $\gamma_{od}.$ For the calculations presented here, the diagonal and off-diagonal relaxation rates are $\gamma_d=0.1\operatorname{1/fs}$ and $\gamma_{od}=0.3\operatorname{1/fs}$, respectively.
As one can check, using dipole approximation and velocity gauge, $\rho$ will always be diagonal in the index $\mathbf{k},$ that is, the dynamics mix only the band indices ("vertical transitions"), but not $\mathbf{k}$'s.

In thermal equilibrium (without external bias), the sample does not radiate, the net current flowing through it is zero. When a laser pulse impinges the crystal, this changes, and the nonzero current density
\begin{equation}
\mathbf{J}=\frac{e}{\mathcal{V}m}\mathrm{Tr}\left[ \rho \mathbf{p}_{kin}\right] =\frac{e}{\mathcal{V}m} \mathrm{Tr} \left[\rho (\mathbf{p}-e\mathbf{A})\right].
\label{Jpure}
\end{equation}
becomes the source of the HHG signal. Note the appearance of the kinetic momentum $\mathbf{p}_{kin}$ here: since it is proportional to the velocity operator (which does not hold for the canonical momentum $\mathbf{p}=-i\hbar \nabla$), its expectation value is proportional to the macroscopic net current density that appears in Maxwell's equation as a source. For linearly polarized excitation, the power spectrum of the only nonzero component of $\mathbf{J}$ is assumed to provide the HHG spectra.

Note that $\mathbf{J}$ is often considered as a sum of "polarization-like" interband and "current-like" intraband components. Since this distinction -- at least in the single electron picture we use here -- is shown to be gauge dependent \cite{FP2017}, we shall not use it in the following. Instead, we consider the gauge independent, entire $\mathbf{J}$ as the source of the HH radiation.

\subsection{Numerical approach}
In order to simplify the calculations, we are going to use a one-dimensional model in the following (see Fig.~\ref{Fig1}). Note that this can be adequate for linearly polarized excitations. We consider two model potentials that have one point in common: both produce a band scheme with a gap of $3.2\operatorname{eV},$ which corresponds to the case of the ZnO crystal. The explicit forms of the two potentials can be written as:
\begin{equation}
U^{(a)}(x)=-U^{(a)}_{0}\sum_{i=1}^{2} \cos^{2}\left[\pi\left(x-x^{(a)}_{i}\right)/\Delta^{(a)}\right],
\label{cospot}
\end{equation}
and
\begin{equation}
U^{(b)}(x)=-U^{(b)}_{shift}+U^{(b)}_{0}\sum_{i=1}^{2} \sinh^{2}\left(x-x^{(b)}_{i}\right),
\label{sinhpot}
\end{equation}
where $U^{(a)}_{0}=25\,\operatorname{eV},$ $\Delta^{(a)}/a=0.15,$ $x^{(a)}_{1}/a=0.3$ and $x^{(a)}_{2}/a=0.607$. The parameters related to $U^{(b)}$  are the following: $U^{(b)}_{shift}=187.722\,\operatorname{eV},$ $U^{(b)}_{0}=4080.925\,\operatorname{eV},$ $x^{(b)}_{1}/a=0.18,$ and $x^{(b)}_{2}/a=0.7$.
Note that apart from the band gap, these potentials correspond to qualitatively different band schemes, thus they can be used to check to what extent our results depend on the particular choice of the model.

Inserting these potentials in the 1D version of Eq.~(\ref{keigen}), we obtain the Bloch states $\Psi_{n,k}(x).$ We use a finite number of $k$ values, which are uniformly distributed in the first Brillouin zone $\left[-\pi/a, \pi/a\right],$ where $a$ is the lattice constant. As a next step, we calculate the matrix elements $\langle n,k|H_{0}|n^{\prime},k^{\prime}\rangle$ and $\langle n,k|p_x|n^{\prime},k^{\prime}\rangle,$ which are both proportional to $\delta_{kk^{\prime}}.$ Using these matrix elements and the initial conditions (\ref{rho0}), the dynamical equation (\ref{vN}) can be integrated by numerical means. In these calculations we assume that the only nonzero component of the vector potential is given by
\begin{equation}
A_{x}(x,t)= \left\{ \begin{array}{lll}
A_{0}\sin^{2}(\pi t/\tau)\sin(\omega_{L}t+\varphi), \mbox{ if}\,\, t\in[0,\tau],\\
0\,\,\mbox{otherwise},
\end{array}\right.,
\end{equation}
where $\tau$ measures the duration of the pulse and $\omega_{L}$ denotes the central frequency of the excitation. In dipole approximation, the vector potential has no spatial dependence. Note that since we are considering relatively long pulses ($\approx$30 optical cycles at intensity FWHM), the results are expected to be almost independent of the carrier-envelope phase, $\varphi$. Therefore we use $\varphi=0$ in the rest of the paper.

When the time-dependent density matrix is obtained, the expectation value of the kinetic momentum can be calculated leading to the function $J(t)$ (the 1D version of Eq.~(\ref{Jpure})), and fast Fourier transform gives to the HHG spectra.

The fact that all transitions are vertical in velocity gauge allows us to calculate the contribution of every value of $k$ to the net current density separately. Although the time evolution of the projectors $|n_{0},k\rangle \langle n_{0},k|$ is independent for different values of $k,$ and, according to (\ref{rho0}), there is no quantum mechanical interference between these states either, their contribution to $J$ has to be added linearly, so these contributions do interfere. Technically, this means that first we have to construct $J(t)$ and then calculate its power spectrum (so it is not the power spectra corresponding to different values of $k$ that has to be added.) In the next section we investigate how these interference phenomena determine the properties of the HHG spectra.

\section{Results}
\label{resultsec}
\subsection{General properties of the HHG spectra}
\label{spectra}
\begin{figure}[tbh]
\includegraphics[width=8.5cm]{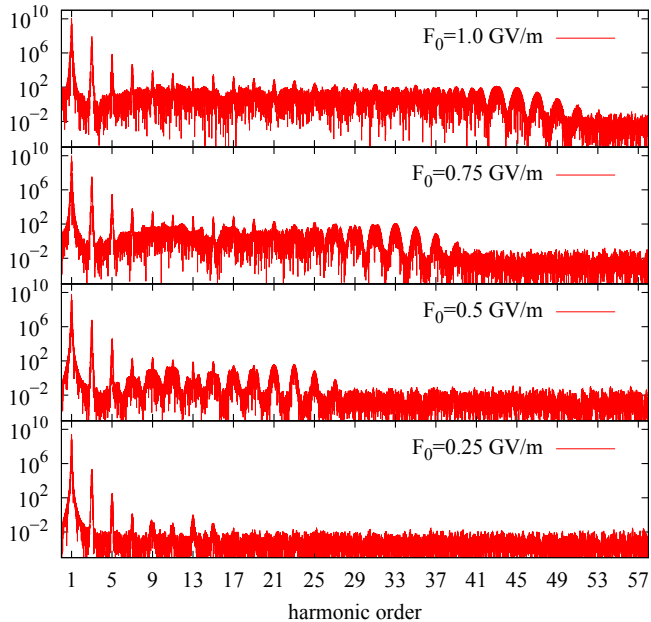}
\caption{Representative HHG spectra for different peak field strengths.  The band scheme of the target is derived using the potential $U^{(a)}.$ Parameters: $\lambda_{L}=\SI{3}{\micro\metre}$ and $\tau=300\operatorname{fs}.$ The frequency on the horizontal axis is given in units of $\omega_{L}=2\pi c/\lambda_{L}$.}
\label{intensityfig}
\end{figure}
The HHG spectra that can be obtained as described above show a weak dependence on the number of the conduction bands that we take into account. According to our experience, for moderate exciting field intensities, a 4 band model (VB+3CBs) is sufficient in the sense that adding more CBs the results practically do not change. Additionally, as expected, the choice of the periodic potential $U$ (as given by Eqs.~(\ref{cospot}) and (\ref{sinhpot})) also influences the resulting spectra. However, we found that the qualitative features of the HHG spectra to be discussed below, are general, i.e., the same for both potentials. Therefore, apart from a single example (see the next subsection) we are not going to compare all the results for $U^{(a)}$ and $U^{(b)}.$

Fig.~\ref{intensityfig} shows representative HHG spectra calculated using $U^{(a)}.$ As we see, the usual structure of the high order harmonic spectra can be identified in this figure. The most dominant peak corresponds to the exciting frequency, there is a plateau region with peaks of approximately the same heights, and a cutoff where the peaks disappear. The difference between the maximum and minimum of the spectra is several orders of magnitude.
Additionally, in this intensity regime, the cutoff scales linearly with the amplitude of the exciting field, in accord with experimental findings \cite{GDiCSADiMR11}. Moreover, as a closer look reveals, peaks corresponding to odd order harmonics are considerably higher than the ones at frequencies that are even integer multiples of the central frequency of the excitation.
Note that for larger excitation intensities, the position of the cutoff is more difficult to identify. Similarly to the case of a two-level system \cite{GCVF2016}, there is no overall drop in the rapidly oscillating spectral curves, the peaks rather tend to merge into the continuum.

Based on the properties collected above, we can conclude that our 1D model can reproduce most of the characteristic properties of the HHG spectra. Now we investigate the physical reasons for the appearance of these properties.

\subsection{Even and odd harmonics}
\label{evenodd}
\begin{figure}[tbh]
\includegraphics[width=8.5cm]{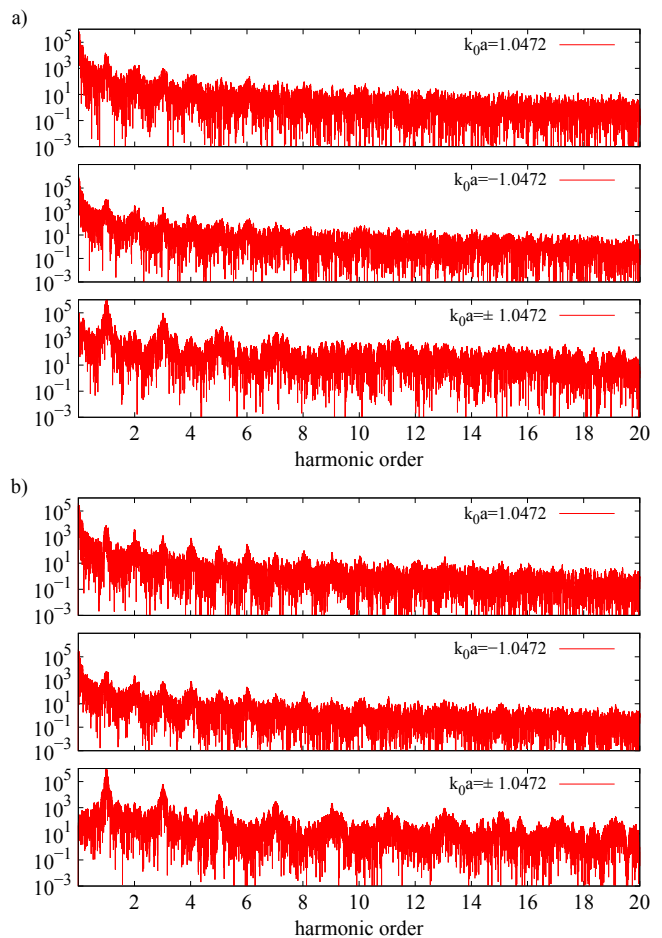}
\caption{The suppression of even order harmonics. Panel $a)[b)]$ corresponds to $U^{(a)} [U^{(b)}].$ In both panels, the contributions of given values of $k$ is shown together with that of $-k$ and their combined effect, for
parameters $F_{0}=4.0\operatorname{GV/m}$, $\lambda_{L}=\SI{3}{\micro\metre}$ and $\tau=300\operatorname{fs}.$}
\label{evenoddfig}
\end{figure}
As we have seen, it is possible to calculate spectra that result from any subset of $k$ values chosen from the first Brillouin zone. In other words, we can artificially restrict the summation over $k$ Eq.~(\ref{rho0}), and investigate how interference of the HH radiation emerging from these states as sources contribute to the net HHG signal.

Focusing on the practical absence of the even order harmonics from the spectra, Fig.~\ref{evenoddfig} provides an explanation. As we can see, HHG signals corresponding to a given value of $k$ contain all harmonics with approximately the same weights. However, as the example of Fig.~\ref{evenoddfig} shows, when we combine the fields resulting from the time evolution of a given $k$ and its opposite, the even harmonics tend to disappear. That is, the fields emerging from the time evolution of states that -- without the external field -- have phase velocities with the same magnitudes but opposite signs, interfere constructively for the odd harmonics, and destructively for the even ones. Additionally, this result does not depend on the particular choice of the model potential $U,$  since it is visible for both $U^{(a)}$ and $U^{(b)}.$

In other words, the origin of the asymmetry between even and odd harmonics for bulk solid state HHG is the interference of the fields emitted by states corresponding to opposite $k$ values.

\subsection{The position of the cutoff}
\label{cutoff}
\begin{figure}[tbh]
\includegraphics[width=8.5cm]{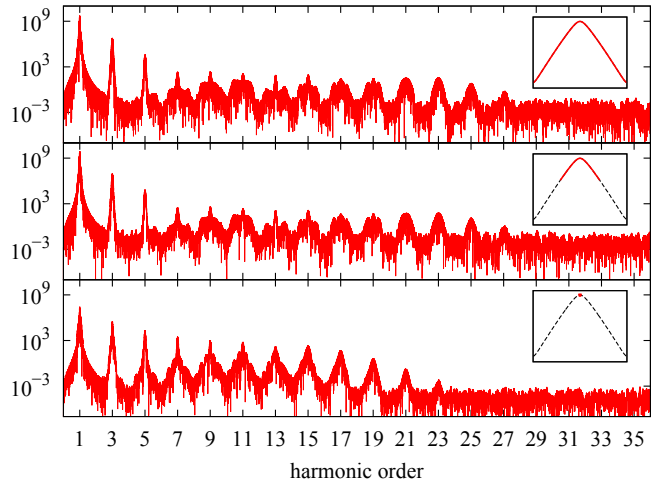}
\caption{Contributions of different parts of the valence band (shown by the inset) to the HHG spectra from potential $U^{(a)}$ represented lattice. Parameters: $F_{0}=0.5\operatorname{GV/m}$, $\lambda_{L}=\SI{3}{\micro\metre}$ and $\tau=300\operatorname{fs}.$ Note that the physical, measurable spectrum is the topmost one, where all initial states from the VB of the first Brillouin zone contribute to the radiation. }
\label{cutoffig}
\end{figure}
For $N$ CBs and a given value of $k,$ we have to calculate the dynamics of an $(N+1)$-level quantum system. For long excitations, Floquet's method \cite{F883} can be applied, and (at least for $N=1,$ see \cite{GGK97}) the cutoff of the HHG spectrum can be analytically estimated. However, in general, these cutoffs will be different for different values of $k$, since the parameters (the $k$-dependent band gap and the matrix elements of the Hamiltonian) will also be different. Additionally, different values of $k$ contribute to the net HHG signal to a different extent. The combined effect of all these contributions result in the HHG spectra that can be detected.

As an example, Fig.~\ref{cutoffig} shows how the net HHG signal and the physically measurable cutoff is being built up as a superposition of the individual contributions from different values of $k$ that cover intervals of increasing sizes around $k=0.$ (The corresponding parts of the valence band are schematically shown by the insets.) Note that, as we saw in the previous subsection, the symmetry of these intervals around zero provides the dominance of the even order HHG peaks in the spectrum over the odd order ones.

At this point it is worth returning to the tendency seen in Fig.~\ref{intensityfig}. The crucial parameter for the Floquet description of HHG for a single few-level system is $A_{0}/\omega_{L}.$ For not too strong, many cycle excitations, all the contribution from different values of $k$ produce a cutoff that is linear in the amplitude of the exciting field, meaning that the net HHG signal has to behave similarly. However, in view of Fig.~\ref{cutoffig} and the previous subsection, it is important to stress that the physical, observable HH radiation stems from the \emph{interference} of fields related to the time evolution of initial states with all values of $k.$

\section{Summary}
\label{summ}
We considered a quantum mechanical model for the generation of high-order harmonics in bulk solids. The crystal was assumed to be initially in thermal equilibrium, with the valence band being populated only. We investigated the contribution of different initial states to the HH radiation and showed that important aspects of the HHG spectra are strongly influenced by the interference of these contributions. According to our calculations, the absence of the even-order harmonics can be explained by the destructive interference of fields corresponding to initial states with opposite crystal momentum $k.$ The position of the cutoff is also shown to originate from interference effects.

\section*{Acknowledgments}
Our   work   was   supported   by   the   Hungarian National Research, Development and Innovation Office under Contract No. NN 107235. Partial support by the ELI-ALPS project is also acknowledged. The ELI-ALPS project (Grants No.  GOP-1.1.1-12/B-2012-000  and  No.  GINOP-2.3.6-15-2015-00001)  is  supported  by  the  European  Union  and  co-financed by the European Regional Development Fund.
Our project was also supported by the Hungarian National Talent Programme under Contract No. NPT-NFT\"{O}-16-0994, and by the European Social Fund under contract EFOP-3.6.2-16-2017-00005.

\end{document}